\documentclass{article}
\usepackage{ismir,amsmath,cite,url}
\usepackage{graphicx}
\usepackage{color}
\usepackage{url}
\usepackage[group-four-digits]{siunitx}
\usepackage{booktabs}

\usepackage{listings}

\title{Piece Identification in Classical Piano Music Without Reference Scores}

\oneauthor
 {Andreas Arzt, Gerhard Widmer}
{Department of Computational Perception, Johannes Kepler University, Linz, Austria \\Austrian Research Institute for Artificial Intelligence (OFAI), Vienna, Austria\\{\tt andreas.arzt@jku.at}}

\begin{document}

\maketitle
\begin{abstract}

In this paper we describe an approach to identify the name of a piece of piano music, based on a short audio excerpt of a performance. 
Given only a description of the pieces in text format (i.e. no score information is provided), a reference database is automatically compiled by acquiring a number of audio representations (performances of the pieces) from internet sources.
These are transcribed, preprocessed, and used to build a reference database via a robust symbolic fingerprinting algorithm, which in turn is used to identify new, incoming queries.
The main challenge is the amount of noise that is introduced into the identification process by the music transcription algorithm and the automatic (but possibly suboptimal) choice of performances to represent a piece in the reference database.
In a number of experiments we show how to improve the identification performance by increasing redundancy in the reference database and by using a preprocessing step to rate the reference performances regarding their suitability as a representation of the pieces in question. As the results show this approach leads to a robust system that is able to identify piano music with high accuracy -- without any need for data annotation or manual data preparation.

\end{abstract}

\section{Introduction}\label{sec:introduction}

Efficient algorithms for content-based audio retrieval enable systems that allow users to browse and explore music collections (see e.g. \cite{grosche:mmp:2012} for an overview). In this context \emph{audio fingerprinting} algorithms which permit the fast identification of an unknown recording (as long as an almost \emph{exact replica} is contained in the reference database) play an important role. For this task there exist highly efficient algorithms that are in everyday commercial use (see e.g., \cite{cano:mmsp:2002,wang:ismir:2003,ramona:icassp:2013,baluja:pr:2008,six:ismir:2014,sonnleitner:tasl:2016}).

However, these algorithms are not able to identify \emph{different performances} of the \emph{same piece} of music, as they are not designed to work in the face of musical variations such as different tempi, expressive timing, differences in instrumentation, ornamentation and other performance aspects. Regarding classical music, the identification of performances that derive from a common musical score is of special interest, as in general there exists a large number of performances of the same piece (and new renditions are performed every day).

This task is generally called \emph{audio matching} (or, mostly in the context of popular music, \emph{cover version identification}, see e.g. \cite{serra:amir:2010}). A common approach to solve this problem is to use an \emph{audio alignment} algorithm. This is computationally expensive, as it basically involves aligning the query snippet with every position within every audio file in the database (see \cite{mueller:ismir:2005}, and \cite{kurth:tasl:2008} for a indexing method that makes the problem more tractable). Furthermore, due to the coarse feature resolution of these algorithms, relatively large query sizes are needed.

As there exist efficient fingerprinting algorithms, it seems natural to try to adapt them to the problem of cover version identification. A first study towards this is presented in \cite{grosche:icassp:2012b}, where the authors focused on the suitability of different low-level features as a basis for fingerprinting algorithms, but neglected the problem of tempo differences between performances. In \cite{arzt:ismir:2012} an extension to a well-known fingerprinting algorithm\cite{wang:ismir:2003} is proposed that makes it invariant to the global tempo. With the help of an \emph{audio transcription} algorithm for piano music (see \cite{boeck:icassp:2012}) a system was built that, given a short audio query, almost instantly returns the corresponding (symbolic) \emph{score} from a reference database -- despite the fact that audio transcription is a very hard problem and thus introduces a lot of noise in the process.

In this paper we show how to use this algorithm in the absence of symbolic scores to identify unknown performances, using a reference database based on other \emph{performances} of the pieces in question. As symbolic scores are often not readily available, this increases the applicability of this algorithm in real life systems. The downside of this approach is that now audio transcription is used for both the data contained in the reference database and for the queries, which introduces even more noise. Furthermore, the transcription algorithm we are using is optimised on piano sounds, which for now limits the proposed system to piano music only.

We are going to describe this approach in the context of a system geared towards fully automatic identification of classical piano music, in the sense that even the creation of the collection of audio recordings, which is needed to perform the identification task, is automated. The motivation for this is to reduce the amount of costly manual annotation to a minimum, and instead facilitate available, albeit noisy, web sources like \textit{YouTube}\footnote{\url{https://www.youtube.com}} or \textit{Soundcloud}\footnote{\url{https://soundcloud.com}}. The main challenge in this setting is the noise introduced into the identification process via multiple processes (automatic retrieval of reference performances, audio transcription of reference performances, and audio transcription of the query). In the paper we will show how to deal with this amount of noise by increasing redundancy in the reference database and by an automatic selection strategy for the reference performances.

The paper is structured as follows. Section \ref{sec:overview} gives an overview of the proposed system. Then, in Section \ref{sec:dataset} the data we are using for our experiments is described. Sections \ref{sec:baseline}, \ref{sec:multiple}, \ref{sec:selecting} and \ref{sec:multiple_queries} describe the core experiments of the paper, showing that our approach is robust enough to cope with the multiple sources of noise and performs well in our experiments. A brief outlook on possible improvements and applications is given in Section \ref{sec:conclusions}.

\section{System Overview}\label{sec:overview}

\begin{figure*}
 \centerline{
 \includegraphics[width=\textwidth]{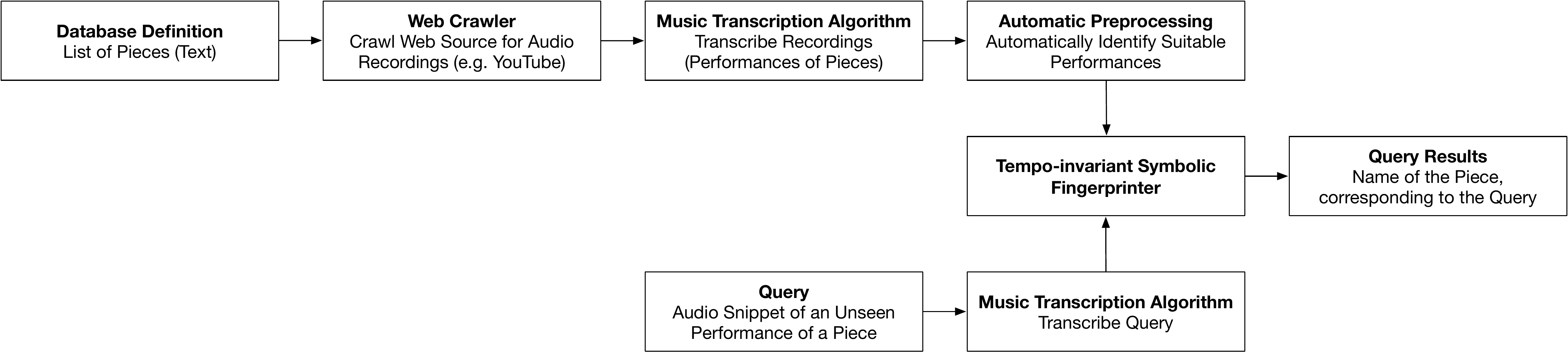}}
 \caption{System Overview}
 \label{fig:system_overview}
\end{figure*}

In this section we are going to describe the piece identification system that will be used throughout the paper. The main goals of the system are 1) to automate the process of compiling a reference database, thus making manual annotations obsolete, and 2) based on this reference database, allow for robust and fast piece identification. Figure \ref{fig:system_overview} depicts how the components interact with each other.

The system is based on a \textbf{Database Definition} file, which is a list of pieces that are to be included in the database. On this list each piece is represented by an ID, the name of the composer and the name of the piece, including identifiers like the opus number (see Figure \ref{fig:excerpt_pieces} for an excerpt of the list). We would like to emphasise once more that this is the only input our system needs (in addition to a source from which the recordings can be retrieved). All the data necessary to perform the identification task is then prepared automatically. This also means that extending the database is as easy as adding a new line to the text file, describing the new piece. The data in this file also defines the granularity of the database. For example, movements of a sonata could be represented as individual pieces or combined as single piece -- for our experiments we took the latter approach. For our proof-of-concept implementation we settled for 339 piano pieces of well-known composers (Mozart, Beethoven, Chopin, Scriabin, and Debussy), which already represents a substantial share of the classical piano music repertoire.

\begin{figure}
\begin{lstlisting}[
    basicstyle=\tiny
]
ID;Composer;Piece
...
17;Mozart;Piano Sonata No. 17 in B-flat major K 570
18;Mozart;Piano Sonata No. 18 in D major K 576
19;Mozart;Fantasy No. 1 with Fugue in C major K 394
20;Mozart;Fantasy No. 2 in C minor, K 396
...
41;Beethoven;Piano Sonata No. 14, Op. 27, No. 2 "Moonlight"
42;Beethoven;Piano Sonata No. 15, Op. 28 "Pastoral"
...
168;Chopin;Mazurka Op. 7 No. 5 in C major
169;Chopin;Nocturne Op. 15 No. 1 in F major
170;Chopin;Nocturne Op. 15 No. 2 in F-sharp major
171;Chopin;Nocturne Op. 15 No. 3 in G minor
...
281;Debussy;L 113, Children's Corner, Doctor Gradus ad Parnassum
282;Debussy;L 113, Children's Corner, Jimbo's Lullaby
...
332;Scriabin;Piano Sonata No. 3, Op. 23
333;Scriabin;Piano Sonata No. 4, Op. 30
...
\end{lstlisting}
 \caption{An excerpt of the file used for collecting the database.}
 \label{fig:excerpt_pieces}
\end{figure}

A \textbf{Web Crawler} takes this list of pieces and retrieves audio recordings of performances of the pieces. In our case we use a simple crawler for \textit{YouTube} (an alternative would be to use \textit{Soundcloud}, amongst others). The queries are constructed by concatenating the name of the composer and the piece, and adding the word ``piano'', to ensure that mainly piano performances are returned.

Next, the collected recordings are fed into a \textbf{Music Transcription Algorithm} that takes the audio files and transcribes them into series of symbolic events. For this step we rely on a well known neural network based method presented in \cite{boeck:icassp:2012}, more specifically the version that is available as part of the \textit{Madmom} library \cite{boeck:acmmultimedia:2016}. As input it takes a series of preprocessed and filtered STFT frames with two different window lengths. The neural network consists of a linear input layer with 324 units, three bidirectional fully connected recurrent hidden layers with 88 units, and a regression output layer with 88 units, which directly represent the {MIDI} pitches. The output of the transcription algorithm is a list of detected musical events, represented by their pitches and start times. For details we refer the reader to  \cite{boeck:icassp:2012}. This algorithm exhibits state of the art results for the task of piano transcription, as was demonstrated at the MIREX 2014\footnote{\url{http://www.music-ir.org/mirex/wiki/2014:MIREX2014_Results}}. Still, polyphonic music transcription is a very hard problem, and thus the output of this transcription algorithm contains a relatively large amount of noise, of which the following components need to be robust to.

The \textbf{Automatic Preprocessing} step is concerned with the question of which of the downloaded recordings for each piece should be used in our fingerprint database. In this paper we discuss three setups: take the top match returned by the web crawler (see Section \ref{sec:baseline}), take the top five / fifteen matches returned by the web crawler (see Section \ref{sec:multiple}), and download 30 recordings for each piece, rank them automatically via comparing them to each other and use the top recordings identified via this approach (see Section \ref{sec:selecting}). This means that in the latter two experiments a single piece is represented by \emph{multiple} recordings, adding redundancy to the reference database.

The transcribed sequences of symbolic event information, i.e. sequences of pairs $(\text{pitch}, \text{onset time})$, are fed to the \textbf{Tempo-invariant Symbolic Fingerprinter}, to build a database of fingerprints that later on can be used to identify queries. The algorithm is used as described in \cite{arzt:ismir:2012}, thus it will be summarised here very briefly. The principle idea of the fingerprinting algorithm is to represent an instance (in this case a transcribed performance, representing a piece) via a large number of local, tempo-invariant fingerprint tokens. These tokens are created based on the pitches of three temporally local note events, together with the ratio of their distances in time. Due to the way they are created, the tokens are invariant to the global tempo, and can be stored in a hash table and efficiently queried for. 

An incoming \textbf{Query} is processed in the same way as above by the \textbf{Music Transcription Algorithm}. The resulting sequence of symbolic events is used to query the \textbf{Tempo-invariant Symbolic Fingerprinter} for matches. To do so, from the query the same kind of fingerprint tokens are computed, and matching tokens are retrieved from the fingerprint database. Finally, in this result set continuous sequences of matching tokens, which are a strong indication that the query matches a specific part of a piece stored in the fingerprint database, are identified (via a fast, histogram based approach).

The \textbf{Query Result} is a list of positions within the reference performances that were inserted into the database (see Table \ref{tab:result_example}). The positions in the result set are ordered by their number of tokens matching the query. As can be seen, the result set is actually more detailed than necessary for our applications scenario, as we are only interested in identifying the respective piece, and not a specific reference performance (or even a position within reference performance). Thus for the experiments in this paper we summarise all occurrences of a piece into one score by summing up the matching scores of all its occurrences in the results set.

\begin{table}
 \begin{center}
 \begin{tabular}{cccc}
  \toprule
  {Piece ID} & {Performance ID} & {Time in Ref.} & {Score}  \\
  \midrule
 1 & 0 & 99 & 351 \\
 1 & 0 & 21 & 292 \\
 1 & 4  & 16 & 109 \\
 1 & 4  & 15 & 36 \\
 1 & 4 & 148 & 36 \\
 1 & 4 & 150 &32 \\
10 & 48 & 368 & 7 \\
 1 & 0 & 239 &7 \\
  \bottomrule
 \end{tabular}
\end{center}
 \caption{An example of a result returned by the fingerprinting algorithm. This query was performed on a database in which multiple reference performances represent a piece of music, hence for the piece with ID 1 results for two performances are returned. The score is the number of matching fingerprint tokens for the given query at the specific time in the reference recording. For our purposes we summarise the results per piece, i.e. the matching score for the piece with ID 1 is 863, and for the piece with ID 10 it is 7.}
 \label{tab:result_example}
\end{table}

\section{Groundtruth Data and Experimental Setup}\label{sec:dataset}

For the experiments presented in this paper, ground truth data, i.e. performances for which the composer and the name of the piece is known, is needed. We are using commercial recordings of a large part of the pieces contained in our database. This includes e.g. Uchida's recordings of the Mozart Sonatas, Brendel's recordings of the Beethoven Sonatas, Chopin recordings by Arrau, Pires and Pollini, and Debussy recordings by Pollini, Thibaudet, Zimerman. We would like to emphasise that to get realistic results, in our experiments we made sure manually that no exact replicas of these performances are contained in the automatically downloaded data that is used to build the reference database later on. In total 370 tracks were selected and assigned manually to the respective pieces (roughly 30 hours of music, or \num{665000} transcribed events). Some of the tracks were assigned to the same piece, as e.g. the movements of the sonatas are typically represented as different audio tracks, but are represented as a single piece in our database.

The experimental setup is as follows. We are going to use the same set of randomly extracted queries for each experiment. We are using three query lengths of 2, 5 and 10 seconds (we only took queries though which had at least 10 transcribed notes, avoiding to e.g. query for silence), and extract for each length ten queries for each ground truth performance (giving a total of \num{3700} queries for each query length). The experiments are based on different strategies to automatically compile the reference database. We start with a simple baseline approach (Section \ref{sec:baseline}) and then gradually improve on it by introducing redundancy and a selection strategy (Sections \ref{sec:multiple} to \ref{sec:multiple_queries}).

As evaluation measure we use the \emph{Recall at Rank k}\footnote{We would like to note that the related measure \emph{Precision at Rank k} is not useful in our experimental setup, as there will only be at most one correct result in the result set.}. This is the percentage of queries which have the correct corresponding piece in the first $k$ retrieval results. In our experiments we look at the recall at ranks 1, 5 and 10. In addition, we also report the \emph{Mean Reciprocal Rank} (MRR).

\begin{equation}
\text{MRR} = \frac{1}{|Q|} \sum_{i=1}^{|Q|} \frac{1}{\text{rank}_i}
\end{equation}

Here, $\text{rank}_i$ refers to the rank position of the correct result for the $i^{th}$ query.

The mean query times (i.e. the mean time it takes to process a single query) given in the tables are based on a desktop computer on a single core\footnote{Intel Core i7 6700K 4 GHz with 32 GB RAM.}. If needed, the computation could easily be sped up by multi-threading the query process.

\section{Baseline Approach}\label{sec:baseline}

The baseline approach is very straightforward. The web crawler is used to download the top result from the web source for each piece on the list. The downloaded audio files are transcribed and then processed by the fingerprinting algorithm to build the reference database, i.e. in the reference database each piece is represented by one performance. Note that due to the automatic process the database can be quite noisy, as some of the pieces might be incomplete (e.g. only a single movement of a piece), represented by more than the actual piece (if e.g. the performance downloaded for the piece also contains other pieces, like a recording of a full concert), or the representation is wrong (if the top result of the web crawler is actually a performance of some other piece).

The generated fingerprint database is queried via the prepared excerpts of the collected ground truth data (see Section \ref{sec:dataset}). The results of this first experiment can be seen in Table \ref{tab:baseline}. As can be seen, already in this scenario and despite the small query sizes the method gives reasonable results. For queries of length ten seconds the algorithm returns the correct name of the piece in close to 50\% of the cases. A closer look at the results though showed that the main problem with this simplistic approach is that, as expected, for many pieces the representation in the database is not correct or incomplete. This problem is tackled in the following sections.

\begin{table}
 \begin{center}
 \begin{tabular}{lccc}
  \toprule
   & \multicolumn{3}{c}{Query Length} \\
   & \SI{2}{\second} & \SI{5}{\second} & \SI{10}{\second} \\
  \midrule
  {Recall at Rank 1} & 0.28  & 0.38  & 0.46 \\
  {Recall at Rank 5}  & 0.34  & 0.45 & 0.54 \\
  {Recall at Rank 10} & 0.35 & 0.47 & 0.55 \\
  {Mean Reciprocal Rank} & 0.30 & 0.41 & 0.48 \\
  Mean Query Time & \SI{0.13}{\second} &  \SI{0.41}{\second} &  \SI{0.92}{\second} \\
  \bottomrule
 \end{tabular}
\end{center}
 \caption{Results of the baseline approach. The results are based on \num{3700} queries for each query length.}
 \label{tab:baseline}
\end{table}

\section{Using Multiple Instances Per Piece}\label{sec:multiple}

A simple way to improve the performance of the system is to increase the redundancy within the reference database. Instead of relying on a single instance (recording) for each piece in the reference base, each piece is represented by multiple recordings. For the first experiment five performances per piece were downloaded using the web crawler. The performances were processed in the same way as for the baseline approach in Section \ref{sec:baseline} above and inserted into the fingerprint database. Then, on this database the same set of queries were performed. As described in Section \ref{sec:overview}, the match score of a piece is computed by summing up the scores of the performances representing the piece in question (also see Table \ref{tab:result_example}).

Table \ref{tab:multiple_inst} shows the results of this experiment. As can be seen, the increased redundancy leads to a substantial increase in identification results, compared to the baseline (see Table \ref{tab:baseline}). The added redundancy increases the chances that for each piece at least one ``good'' performance (in the sense of corresponding to the piece and relatively easy to transcribe) is contained in the reference database, and thus mitigates the problems caused by noise, at least to some extent.

For an additional experiment we increased the number of performances to fifteen per piece. These results are shown in Table  \ref{tab:multiple_inst_15}. This improved the results even further. The downside of adding more instances to the fingerprint database is a significant increase in computation time.

\begin{table}
 \begin{center}
 \begin{tabular}{lccc}
  \toprule
   & \multicolumn{3}{c}{Query Length} \\
   & \SI{2}{\second} & \SI{5}{\second} & \SI{10}{\second} \\
  \midrule
  Recall at Rank 1 & 0.58 & 0.69 & 0.74  \\
  Recall at Rank 5  & 0.72  & 0.84 & 0.90 \\
  Recall at Rank 10  & 0.74 & 0.86 & 0.92 \\
  {Mean Reciprocal Rank} & 0.64 & 0.77 & 0.84  \\
    Mean Query Time & \SI{0.34}{\second} &  \SI{0.81}{\second} &  \SI{2.49}{\second} \\
  \bottomrule
 \end{tabular}
\end{center}
 \caption{Results on the reference database based on multiple recordings (the top five results according to the web source) to represent each piece. The results are based on \num{3700} queries for each query length.}
 \label{tab:multiple_inst}
\end{table}

\begin{table}
 \begin{center}
 \begin{tabular}{lccc}
  \toprule
   & \multicolumn{3}{c}{Query Length} \\
   & \SI{2}{\second} & \SI{5}{\second} & \SI{10}{\second} \\
  \midrule
  Recall at Rank 1 & 0.76 & 0.87 & 0.91  \\
  Recall at Rank 5  & 0.84  & 0.94 & 0.97 \\
  Recall at Rank 10  & 0.86 & 0.95 & 0.98 \\
  {Mean Reciprocal Rank} & 0.80 & 0.90 & 0.94  \\
    Mean Query Time & \SI{0.82}{\second} &  \SI{2.85}{\second} &  \SI{6.08}{\second} \\
  \bottomrule
 \end{tabular}
\end{center}
 \caption{Results on the reference database based on multiple recordings (the top fifteen results according to the web source) to represent each piece. The results are based on \num{3700} queries for each query length.}
 \label{tab:multiple_inst_15}
\end{table}

\section{Automatically Selecting Suitable Representations}\label{sec:selecting}

\begin{table}
 \begin{center}
 \begin{tabular}{lccc}
  \toprule
   & \multicolumn{3}{c}{Query Length} \\
   & \SI{2}{\second} & \SI{5}{\second} & \SI{10}{\second} \\
  \midrule
  Recall at Rank 1 & 0.54 & 0.68  & 0.74  \\
  Recall at Rank 5  &  0.63 & 0.76 & 0.83 \\
  Recall at Rank 10  & 0.64 & 0.78 & 0.85 \\
  {Mean Reciprocal Rank} & 0.58 & 0.72  & 0.78  \\
    Mean Query Runtime & \SI{0.14}{\second} &  \SI{0.47}{\second} &  \SI{0.97}{\second} \\
  \bottomrule
 \end{tabular}
\end{center}
 \caption{Results on the reference database based on the top recording selected via the proposed strategy to represent each piece. The results are based on \num{3700} queries for each query length.}
 \label{tab:selecting_single_inst}
\end{table}

\begin{table}
 \begin{center}
 \begin{tabular}{lccc}
  \toprule
   & \multicolumn{3}{c}{Query Length} \\
   & \SI{2}{\second} & \SI{5}{\second} & \SI{10}{\second} \\
  \midrule
  Recall at Rank 1 & 0.72 & 0.85 &  0.89 \\
  Recall at Rank 5  & 0.82  & 0.92 & 0.96 \\
  Recall at Rank 10  & 0.84 & 0.93 & 0.97 \\
  {Mean Reciprocal Rank} & 0.77 & 0.88 & 0.92  \\
    Mean Query Time & \SI{0.49}{\second} &  \SI{1.71}{\second} &  \SI{3.83}{\second} \\
  \bottomrule
 \end{tabular}
\end{center}
 \caption{Results on the reference database based on multiple recordings (top five recordings selected via the proposed strategy) to represent each piece. The results are based on \num{3700} queries for each query length.}
 \label{tab:selecting_inst}
\end{table}

A closer look at the results so far shows that increasing the redundancy in the reference database indeed leads to better results, but also increases the computation time. The main problem with our approach is that in addition to useful data, the process also adds a lot of extra noise to the fingerprint database. The web crawler returns a considerable number of performances of the wrong piece, performances played on a different instrument, and performances recorded in very bad quality. This kind of data increases the runtime and decreases the identification accuracy. In this section we present a method for identifying performances in a given a set of candidates for a piece that most probably are related to the piece in question, which also enables us to discard performances that most probably are noise. In this way we try to reduce the number of stored fingerprint tokens, which generally decreases the computation time, while still achieving good identification performance.

Thus, for each piece we perform the following process to select appropriate representations. First, 30 recordings are downloaded via the web crawler. With a high probability at least some of these are actually piano performances of the piece we are looking for, while the others might have nothing in common. The idea now is to find a homogenous group within this set of candidates. To identify performances which are part of this group, we again employ the symbolic fingerprinting process, but limited to the set of candidate performances. To do so, the performances are transcribed and inserted into a new fingerprint database. 

The intuition is that for a query extracted from the same set of candidate performances (that actually matches the piece), the fingerprinter will likely return three kinds of results. Firstly, the top result will be the performance the query was taken from. This is a perfect fit for all tokens, which results in the maximum score. Secondly, a number of other performances will probably also have a high score, identifying them as being based on the same piece and as being transcribed in sufficient quality. Thirdly, performances that actually belong to a different piece, or which are transcribed poorly, will score very low.

Based on these observations, we designed the process of ranking the performances regarding their suitability to represent the piece in question as follows. For each of the performances ten queries are randomly extracted (for our experiments we used a query length of ten seconds) and processed by the fingerprinting algorithm. As in all other experiments, the results are summarised on the performance level (i.e. match scores of positions within the same performance are summed up). Then, for each result the score of the top match (i.e. of the performance the query stems from) is stored, this performance is removed from the result set, and the remaining matching scores are normalised by dividing by the top match score. The reasoning behind this is that the absolute scores depend on the particulars of the query (foremost the length in the sense of the number of notes, but also e.g. if the part in question is normally played in a steady tempo or is subject to expressive tempo changes, which makes it harder to detect and leads to a lower score).

This results in 300 preprocessed and normalised result sets. The suitability of a performance to represent the piece in question is computed by summing up all the scores of all its occurrences in the result sets. The higher this value is for a performance, the more it has in common with the other performances assigned to the piece in question.

Based on this ranking we repeat experiments from Sections \ref{sec:baseline} and \ref{sec:multiple}, but this time for each piece we select the top one or top five performances, respectively, according to the computed rank within the candidate set for each piece. The results are shown in Tables \ref{tab:selecting_single_inst} and \ref{tab:selecting_inst}, which should be compared to Tables \ref{tab:baseline} and \ref{tab:multiple_inst}, respectively. As can be seen the selection strategy increases the identification performance for both scenarios and for all query lengths. 

A comparison of Tables \ref{tab:selecting_inst} and \ref{tab:multiple_inst_15} shows that by using the proposed selection strategy a lower number of performances (5 versus 15) is sufficient to achieve comparable identification accuracy. The decreased number of tokens also results in roughly half the computation time.

The runtime actually depends on a number of factors, most importantly the size of the fingerprint database. But of similar influence is the actual number of tokens that are returned by the fingerprint database for a specific query. The reason is that each of these tokens has to be processed individually to come up with the matching score. This also means that queries for pieces which are represented in the database by a large number of performances will actually take \emph{longer} to compute -- a further argument in favour of the selection strategy presented in this section.


\section{Using Multiple Queries Per Performance}\label{sec:multiple_queries}

So far the assumption was that we only have access to a single short query of two to ten seconds. If instead we have access to a full recording, just querying for one short query would be a suboptimal approach. Thus, we tried an additional query strategy on the reference database based on the performance selection strategy from Section \ref{sec:selecting} above.

A standard approach for processing long queries (in this case a whole performance) would be to apply shingling \cite{casey:ismir:2006,grosche:icassp:2012,arzt:ismir:2014}, i.e. splitting longer queries into shorter, overlapping ones and track the results of these sub-queries over time. Here, as proof of concept we use an even simpler method: we select ten random queries from the piece we want to identify, process them individually and sum up the results. This can be seen as adding redundancy (relying on multiple queries instead of a single one) on the query side. We perform this experiment on the reference database based on the top five selected recordings via the proposed strategy. The results are shown in Table \ref{tab:multiple_queries}. As can be seen this again considerably improves the results, and we are getting very close to 100\%. The main cause for this is that the retrieval precision heavily depends on the quality of the transcription. Some parts of a performance are much harder to transcribe than others (e.g. heavily polyphonic parts with a lot of sustain pedal, which are difficult to transcribe correctly). Using multiple queries, randomly distributed over the whole performance, increases the chances that at least some parts are transcribed in good quality, and that together these queries enable high retrieval accuracy.

Finally, we had a closer look at the few performances that were still misclassified and identified two problems. Our approach does not take care of the problem of recordings of full concerts. If included in the reference database for multiple pieces, these will lead to misclassifications. Furthermore, for some pieces only a small number of performances exists, which causes the crawler to return ``similar'' but wrong performances (e.g. performances of other pieces of the same composer). We sketch a possible solution to these problems in Section \ref{sec:conclusions} below.

\begin{table}
 \begin{center}
 \begin{tabular}{lccc}
  \toprule
   & \multicolumn{3}{c}{Querylength} \\
   & \SI{2}{\second} & \SI{5}{\second} & \SI{10}{\second} \\
  \midrule
  Recall at Rank 1 & 0.92 & 0.95   & 0.95  \\
  Recall at Rank 5  & 0.98 & 0.99 & 0.99 \\
  Recall at Rank 10 & 0.99 & 1 & 1 \\
  {Mean Reciprocal Rank} & 0.94 & 0.97 & 0.97 \\
    Mean Query Time & \SI{0.49}{\second} &  \SI{1.71}{\second} &  \SI{3.83}{\second} \\
  \bottomrule
 \end{tabular}
\end{center}
 \caption{Results for querying for a whole performance via ten random small queries with ten seconds each. The results are based on \num{3700} queries for each query length.}
 \label{tab:multiple_queries}
\end{table}

\section{Conclusions and Future Work}\label{sec:conclusions}

In this paper we presented an approach towards piece identification for performances of piano music, based on an automatically compiled reference database using web sources. It is shown that the symbolic fingerprinting method is robust enough to deal with the noise introduced by the transcription algorithms and allows for fast querying in the symbolic domain. Furthermore, increasing the redundancy by using multiple performances to represent a single piece, especially using the proposed selection strategy, largely alleviates the problem of noise introduced by the automatic compilation of the reference database. Additionally, this increases the robustness of the identification process via the fingerprinting algorithm, as 'problematic' sections (e.g. regarding the transcription process) are represented multiple times, thus increasing the chances that the parts in question are well covered by the reference database.

There exist a number of possible improvements regarding the automatic selection of performances for a piece. In our implementation the focus is on increasing the homogeneity within the group of performances for a piece by comparing them to each other. An additional option is to analyse matches on the full reference database and try to find out which performances match well to multiple pieces and exclude them (as they cover multiple songs or were mistakenly assigned to multiple pieces by the crawler).

We are currently in the process of collecting a much larger collection of classical piano music. This dataset will contain a few thousand pieces, covering a large part of the classical piano repertoire\footnote{The reference database is of course compiled automatically (based on the list of pieces), but the preparation of the ground truth for the experiments is a time consuming, manual process.}. On this dataset we are going to conduct experiments regarding the scalability of our approach in terms of runtime and retrieval accuracy.

In the future, we will also investigate the usefulness of the presented approach for non-classical piano music. Preliminary experiments have shown that this is a much harder task, as compared to classical piano music the pieces are not as strictly defined via a detailed score (e.g. popular songs and jazz standards are mostly described via lead sheets). Thus, performances of the same piece differ more heavily than in classical music. Of course we would also like to lift the restriction to piano music and try our method on other genres, but thus far general music transcription is not robust enough to be used with our approach. Hopefully this will change in the future.

Finally, regarding real-world applications, an automatic method to determine which pieces are well covered by the database, and which ones would benefit from manual intervention, would be desirable. This would help to quickly build a reference database which already covers most pieces well, and then to manually add additional references (based on performances, or even on symbolic score data) for pieces the identification algorithm struggles with.

\section{Acknowledgements}

This work is supported by the European Research Council (ERC Grant Agreement 670035, project CON ESPRESSIONE).

\bibliography{ISMIRArzt}

\end{document}